\documentclass[runningheads]{llncs}
\usepackage{hyperref}
\usepackage{graphicx} 
\usepackage{subfig}
\usepackage{booktabs}
\usepackage{amsmath}
\usepackage{tikz}
\usepackage{cite}
\usepackage{threeparttable}
\usepackage{comment}
\usepackage{xcolor}
\usepackage{multirow}
\usepackage{ifthen}
\usepackage{float}
\usepackage{changepage}
\usepackage{listings}
\usepackage{array}

\newcolumntype{M}[1]{>{\centering\arraybackslash}m{#1}}

\newcommand{\marcom}[2]{%
  \ifthenelse{\equal{#1}{EG}}{ 
    {\noindent\footnotesize\tt\textcolor{red}{\fbox{#1} #2}} 
  }{ 
    \ifthenelse{\equal{#1}{AF}}{ 
      {\noindent\footnotesize\tt\textcolor{green}{\fbox{#1} #2}} 
    }{ 
      \ifthenelse{\equal{#1}{RC}}{ 
        {\noindent\footnotesize\tt\textcolor{brown}{\fbox{#1} #2}} 
      }{ 
        {\noindent\footnotesize\tt\textcolor{blue}{\fbox{#1} #2}} 
      }
    }
  }
}

\usetikzlibrary{positioning, shapes.geometric, arrows.meta, fit, calc, decorations.pathreplacing}

\usetikzlibrary{positioning, shapes.geometric, arrows.meta, fit, calc}

\tikzset{
  process/.style={rectangle, draw, minimum width=3cm, minimum height=1cm, align=center},
  decision/.style={diamond, draw, minimum width=3cm, minimum height=1cm, align=center},
  database/.style={cylinder, draw, minimum height=1cm, aspect=0.5, align=center},
  io/.style={trapezium, trapezium left angle=60, trapezium right angle=120, draw, align=center},
  line/.style={draw, -{Latex[]}},
  note/.style={draw=none, align=center, font=\itshape}
}

\let\oldthebibliography\thebibliography
\renewcommand{\thebibliography}[1]{
  \oldthebibliography{#1}
  \setlength{\itemsep}{0pt}
  \setlength{\parskip}{0pt}
}
\begin{document}

\mainmatter        
\title{UK Finfluencers on TikTok: A Longitudinal Analysis of Content, Engagement, and Disclaimer Practices
}
\titlerunning{Finance and UK Finfluencers}


\author{Essam Ghadafi\inst{1} \and Panagiotis Andriotis\inst{2}}
\authorrunning{Essam Ghadafi and Panagiotis Andriotis} 


\institute{School of Computing, Newcastle University, Newcastle upon Tyne, UK\\
\email{essam.ghadafi@newcastle.ac.uk}
\and
Department of Computer Science, University of Birmingham, Birmingham, UK \\
\email{p.andriotis@bham.ac.uk}
}

\maketitle

\begin{abstract}

The rise of social media financial influencers (finfluencers) has transformed how financial information is disseminated to broad and often inexperienced audiences. While these creators may contribute to financial literacy, concerns remain regarding the reliability of their content and the adequacy of risk disclosures.

Using data collected through TikTok’s Research API, we analyze UK finfluencer content, engagement dynamics, disclaimer practices, audience sentiment, and network structure. The primary dataset comprises 13,215 videos and 104,097 comments posted by 71 UK-based finfluencers between April and September 2024, while a follow-up dataset covering October 2025 to March 2026 enables longitudinal analysis of disclaimer practices, engagement trends, and hashtag usage.

Using topic modeling, we identify four dominant themes: Entrepreneurship \& Side Hustles, Property Investing, Active Trading, and Saving \& Budgeting. Sentiment analysis of audience comments reveals predominantly neutral-to-positive responses, while engagement analysis shows only a negligible association between video duration and engagement rate. Social network analysis indicates a collaborative ecosystem in which mid-tier finfluencers frequently act as bridges between creator groups. Explicit disclaimers and risk-related language remain relatively uncommon overall and are concentrated primarily in trading-related content.

The findings highlight challenges related to financial transparency and disclosure practices within short-form financial content ecosystems. We discuss implications for consumer protection and the design of clearer and more standardized financial risk disclosures on social media platforms.

\end{abstract}
\section{Introduction}
\label{sec:intro}
The rise of social media has transformed the dissemination of financial information, with so-called financial influencers (or for short finfluencers) playing an increasingly significant role in shaping public investment behaviour. Platforms such as TikTok and Instagram have become key sources of financial content, where influencers promote investment strategies, trading platforms, and cryptocurrency ventures to wide audiences, often without proper regulatory oversight \cite{young2023,tiktokinsta2024}. While these influencers have the potential to democratise access to financial knowledge, concerns over misinformation, misleading promotions, and the risk of financial harm have drawn growing scrutiny from regulators.

In response to these challenges, the UK's Financial Conduct Authority (FCA) has strengthened its oversight of financial promotions on social media, emphasising that all investment-related content must be fair, clear, and not misleading. Recent enforcement actions, including interviews under caution and legal proceedings against finfluencers, signal a move towards stricter regulation \cite{FCA2024}. This paper analyzes finfluencer content, disclaimers, topics, and audience sentiment to assess how financial advice, marketing, and regulation interact in digital spaces.

TikTok remains one of the world's most popular social media platforms, recording 773 million downloads in 2024~\cite{businessofapps2024}.
Previous research has examined influencer marketing effectiveness on TikTok, often focusing on entertainment-oriented traits such as humour and originality (e.g., \cite{BARTA2023}), as well as the platform’s structural impact on creative labour and content commodification \cite{Hayes2024}. Our study examines UK TikTok finfluencers, analysing their content, engagement, audience interactions, and disclosure practices to better understand financial communication and consumer protection on the platform.
Given TikTok's popularity among youth \cite{2020Tiktokpopular}, examining finfluencers, whose financial advice may lead to harm or regulatory breaches, is vital for safeguarding financial well-being and ensuring proper oversight of social media content.

A preliminary preprint version of this study~\cite{gha25uk} analyzed only the April--September 2024 observation period and did not include the follow-up analysis or revised disclaimer framework adopted here.

\subsection{Research Questions}
This study investigates the following research questions:

\begin{itemize}

\item \textbf{RQ1:}
To what extent are explicit disclaimers, weak disclaimers, and risk-oriented language present in UK finfluencer content?

\item \textbf{RQ2:}
 What engagement and audience interaction patterns are associated with videos containing disclaimer-related language?
\item \textbf{RQ3:}
How do disclosure practices vary across different financial content themes?

\item \textbf{RQ4:}
What structural patterns characterize the UK finfluencer follower network, and how are interactions distributed across influencer tiers or thematic categories?
\end{itemize}
\subsection{Our Contributions}

This paper makes the following contributions:

\begin{itemize}
\item 
We construct and analyze a longitudinal dataset of UK TikTok finfluencers comprising 21,781 videos and 104,097 comments across two observation periods (April--September 2024 and October 2025--March 2026), enabling longitudinal analysis of disclosure practices, engagement, and hashtag trends.

\item We propose a rule-based framework for detecting explicit disclaimers, weak disclaimers, and risk-related language in finfluencer content using keyword matching, contextual filtering, and regular expressions.

\item We conduct an empirical analysis of engagement dynamics, disclosure practices, sentiment, hashtags, and thematic content within the UK finfluencer ecosystem.

\item We analyze follower-network structure using centrality and connectivity measures, identifying bridging roles and interaction patterns across influencer tiers.

\item We provide longitudinal insights into the evolution of finfluencer behavior,
including evidence of evolving hashtag usage and changing content patterns over time.

\end{itemize}

\section{Related Work}
\label{sec:relatedwork}
Promotions and giveaways are commonly used by influencers to grow their follower base, but similar tactics are also exploited by cybercriminals to attract victims and promote illicit content~\cite{2025darkgram}. In cryptocurrency ecosystems, investment scams remain among the most widespread forms of fraud~\cite{web25}, often involving deceptive websites that promise unrealistic returns~\cite{web25}. Prior studies additionally show that influential online figures can significantly amplify attention toward investment opportunities and speculative assets~\cite{nft2024}. Existing research has documented a broad range of cryptocurrency-related frauds, including pump-and-dump schemes, phishing attacks, fake crypto services, NFT scams, and fraudulent ICOs~\cite{imc24}. These findings highlight the importance of examining the credibility, motivations, and transparency practices of financial influencers operating on social media platforms.

Recent studies emphasize the growing role of social networks in shaping financial behavior, particularly among Millennial and Gen Z populations~\cite{cao2020social,Baird2023}. Social media platforms are increasingly used for financial education, peer-to-peer payments, crowdfunding, social commerce, and investment-related discussions~\cite{Baird2023}. At the same time, concerns have emerged regarding financial misinformation and fraud, with reports suggesting that a substantial proportion of young adults obtain financial advice directly from social media~\cite{sortlist2022}. Prior work additionally shows that teens are increasingly exposed to cryptocurrency-related content and speculative investment narratives through online communities and influencer-driven content~\cite{Bouma-Simsetal2024}.

Several studies investigate the influence of finfluencers on investor behavior and financial decision making. Researchers propose finfluencer quality indicators and scoring approaches to distinguish trustworthy influencers from misleading or manipulative actors~\cite{2024beware}. Other work demonstrates that finfluencers can significantly influence retail investors' trading behavior, portfolio decisions, investor attention, and market activity~\cite{2024retail,impact25,2025investors}. However, researchers also question the expertise and credibility of many finfluencers, particularly those promoting speculative or high-risk investments~\cite{singh2024rise}. Existing studies suggest that recommendations promoted by finfluencers are frequently associated with poor post-recommendation performance and elevated risks for investors~\cite{consu2025,2025crypto}.

Researchers have examined factors influencing investor attitudes toward finfluencer content, highlighting the roles of expertise, credibility, transparency, and content quality in shaping financial decision making~\cite{decision2025,fin2024}. Other studies explore how authenticity, disclosure practices, and presentation characteristics (e.g., accreditation, gender, and race) influence follower perceptions and engagement~\cite{real2025,young2023}. The legal implications of finfluencer disclaimers have also received increasing attention, particularly regarding transparency and the management of consumer expectations~\cite{2023legal}. Finally, Hayes and Ben-Shmuel~\cite{Hayes2024} emphasize the growing societal influence of finfluencers and the need for continued research into their disclosure practices and associated risks.

While prior work has explored the influence of finfluencers on investment behavior, financial literacy, and market dynamics, limited research has examined longitudinal disclosure practices, engagement characteristics, audience interactions, and network relationships of TikTok-based finfluencers within a unified empirical framework. Our work addresses this gap through a large-scale longitudinal analysis of UK TikTok finfluencers, focusing on engagement patterns, disclaimer usage, audience sentiment, and network structure.

\section{Methodology}
\label{sec:methodology}
This section describes our study methodology.
\paragraph{\bf Ethical Considerations.}
\label{sec:ethicalapproval}
Prior to data collection, ethical approval was obtained from the authors' universities and funding body. Data was collected using TikTok's official Research API~\cite{TikTokResearchAPI} in accordance with its terms of service and relevant ethical guidelines.
\paragraph{\bf Data Collection.}
\label{sec:Datacollected}
To classify finfluencers, we collected 10,000 TikTok videos for each month from April to September 2024 using the TikTok Research API. We searched for UK TikTok videos containing finance-related keywords and hashtags associated with investing, budgeting, cryptocurrency, passive income, financial literacy, and wealth-building themes. The complete keyword list is as follows: 
\textquotedblleft budgeting\textquotedblright, \textquotedblleft crypto\textquotedblright, \textquotedblleft cryptocurrency\textquotedblright, \textquotedblleft debtfree\textquotedblright, \textquotedblleft debtpayoff\textquotedblright, \textquotedblleft finance\textquotedblright, 
\textquotedblleft financetiktok\textquotedblright, \textquotedblleft financialeducation\textquotedblright, \textquotedblleft financialfreedom\textquotedblright, \textquotedblleft financialliteracy\textquotedblright, 
\textquotedblleft fintok\textquotedblright, \textquotedblleft investing\textquotedblright, \textquotedblleft investingtips\textquotedblright, \textquotedblleft money\textquotedblright, \textquotedblleft moneygamemoneymindset\textquotedblright, 
\textquotedblleft moneymanagement\textquotedblright, \textquotedblleft moneytok\textquotedblright, \textquotedblleft passiveincome\textquotedblright, \textquotedblleft personalfinance\textquotedblright, 
\textquotedblleft realestateinvesting\textquotedblright, \textquotedblleft retirementplanning\textquotedblright, \textquotedblleft richmindset\textquotedblright, \textquotedblleft savemoney\textquotedblright, 
\textquotedblleft stockmarket\textquotedblright, \textquotedblleft stocks\textquotedblright, \textquotedblleft wealth\textquotedblright, \textquotedblleft wealthbuilding\textquotedblright.

We identified users who accumulated at least 50 comments across their videos and posted at least one video in each month of the April--September 2024 observation period. This threshold excluded low-activity accounts, yielding 71 finfluencers for analysis. A six-month observation period was chosen to capture sustained creator activity. From these accounts, we collected 13,215 videos, including 4,650 containing spoken transcripts in the \texttt{voice\_to\_text} field, together with 104,097 associated comments.

To assess whether observed patterns persisted over time, we collected a follow-up dataset from the same finfluencer cohort covering October 2025 to March 2026. During this period, 12 of the original 71 accounts were no longer accessible through TikTok or the API, resulting in a final cohort of 59 active accounts. The follow-up dataset contains 8,565 videos and was used for analyses of disclaimer usage, engagement, and hashtag trends. Comment-level analyses were restricted to the April--September 2024 dataset because follow-up comments were not collected.

Due to TikTok Research API governance requirements, the raw dataset and identifiers are not publicly redistributed.

\section{Social Graph and Profile Characteristics of Finfluencers}
\label{sec:findata}
This section addresses RQ4 through analysis of network structure and connectivity patterns among finfluencers.

Beyond thematic content, we examined credibility signals and disclosure practices. Only two finfluencers included a bio disclaimer stating that their content did not constitute financial advice, while three accounts were verified. The limited use of profile-level disclaimers suggests relatively low emphasis on transparency and risk communication despite the financial nature of the content.

\begin{figure}[htb]
    \centering
    \includegraphics[width=1\textwidth, height=6cm]{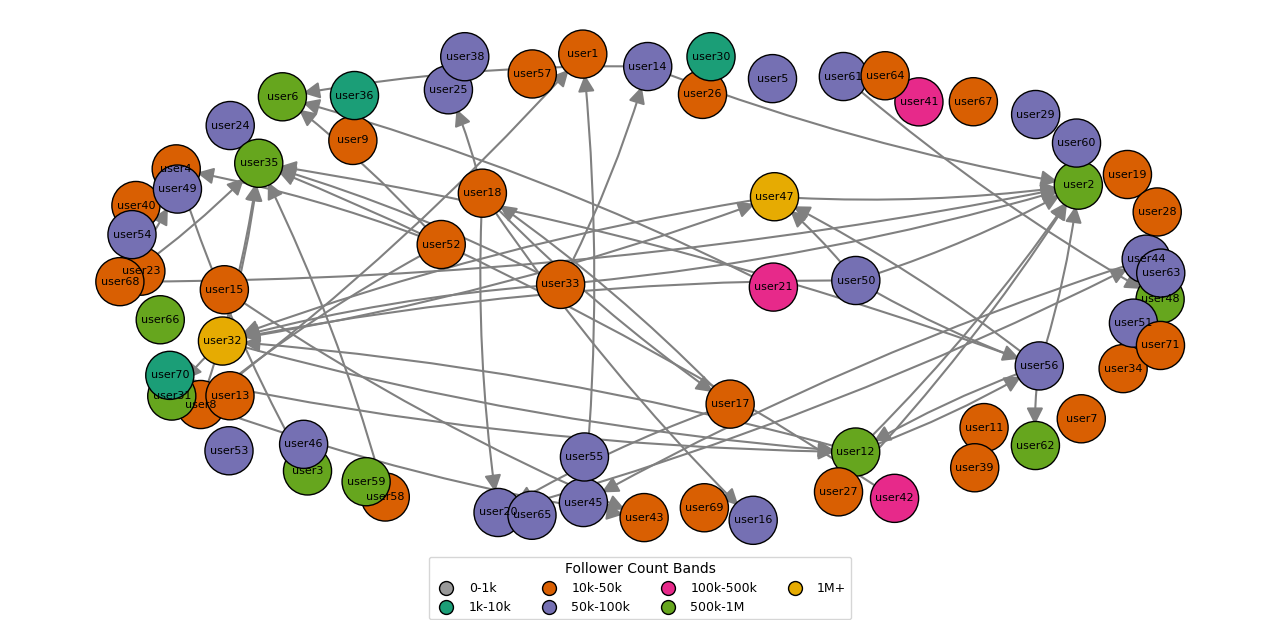}
    \caption{Directed follower network among finfluencers. }
    \label{fig:follower_graph}
\end{figure}

To investigate relationships between finfluencers, we constructed a directed social graph based on mutual interactions. Figure~\ref{fig:follower_graph} illustrates follower relationships between anonymized users represented using pseudonyms (e.g., \texttt{useri}) and follower-count ranges.

We computed closeness and betweenness centrality to identify structurally influential accounts within the network. \texttt{user2} and \texttt{user35} exhibited the highest closeness centrality, indicating efficient connectivity to other users, while \texttt{user32} and \texttt{user12} showed the highest betweenness centrality, suggesting important bridging roles between communities.

Overall, the network exhibited a relatively well-connected structure with several users acting as central hubs. Influencers in the 10k--100k follower ranges appeared most central, suggesting that mid-tier creators play an important role in maintaining connectivity across the UK finfluencer ecosystem.



\section{Video Data Analysis}
\label{sec:Videos}
In this section, we analyse video-level data. Unless otherwise stated, topic modeling and comment-level sentiment analysis use the April--September 2024 dataset, while the follow-up dataset is used for longitudinal analyses of engagement, disclaimer practices, and hashtag trends. This section addresses RQ1--RQ3.
\subsection{Video Duration vs.~Engagement }
\begin{figure}[htb]
    \centering
    \includegraphics[width=0.98\textwidth,height=6cm,keepaspectratio]{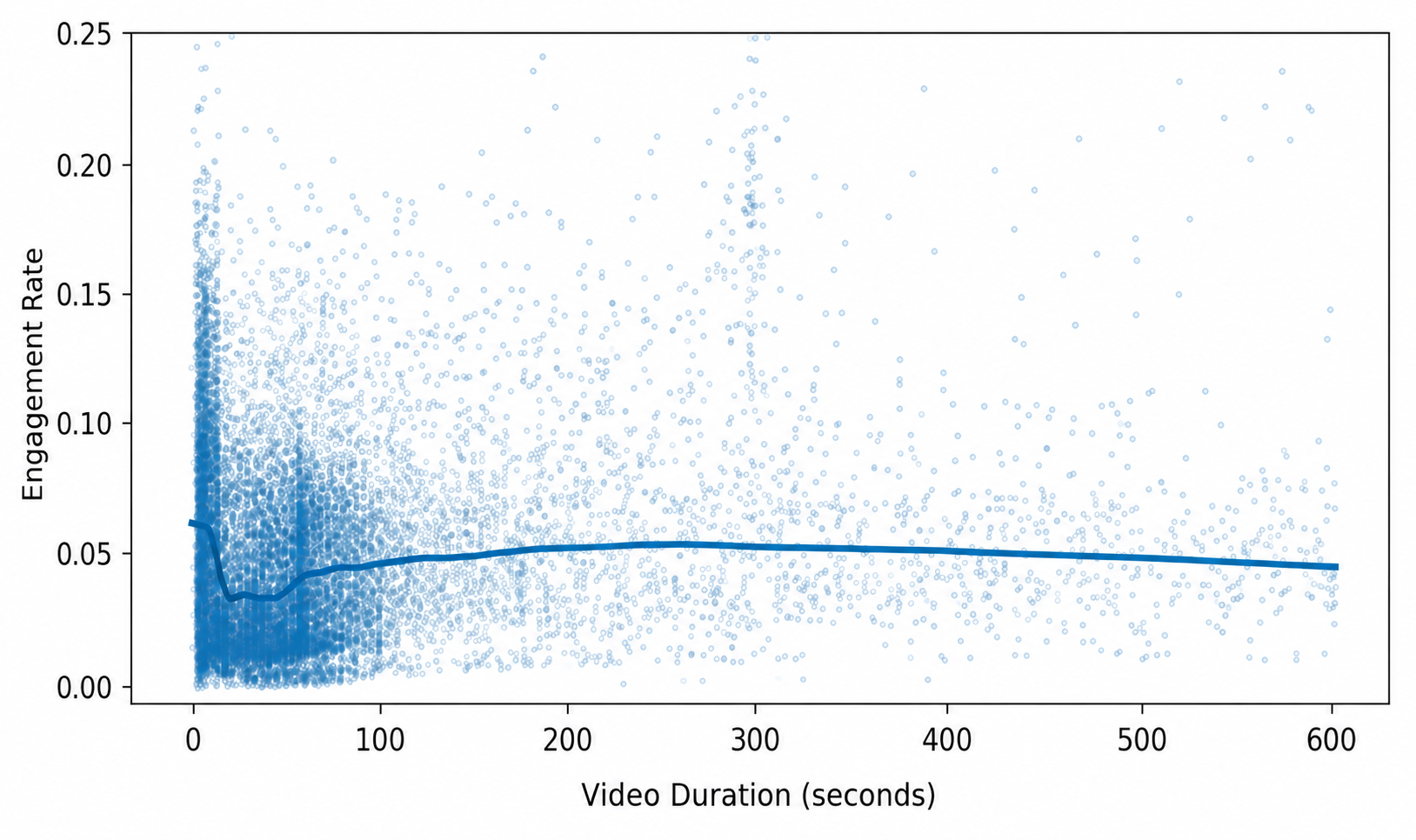}
    \caption{Video duration vs.~engagement rate across finfluencer videos.}
    \label{fig:video_engagement}
\end{figure}

This analysis contributes to RQ2 by examining engagement characteristics within finfluencer videos.
Engagement rate was computed as the ratio of likes, shares, and comments to total views.
Figure \ref{fig:video_engagement} illustrates the relationship between video duration and engagement rate across videos from both observation periods. Most videos were relatively short in duration, with a median duration of 47 seconds and an interquartile range between 19 and 80 seconds. The figure also highlights substantial variability in engagement across all duration ranges.

To reduce the influence of anomalous records and atypical long-form uploads, videos with missing metadata, zero views, zero duration, or durations exceeding 600 seconds were excluded from the analysis. Due to the non-normal distribution of engagement metrics, Spearman's rank correlation coefficient was used to assess the relationship between video duration and engagement.

The statistical analysis revealed only a negligible overall association between duration and engagement ($\rho = 0.015$, $p < 0.05$).
While the correlation was statistically significant, the observed effect size was negligible, indicating that video duration alone was not meaningfully associated with engagement outcomes within the analyzed finfluencer dataset. Engagement is therefore likely influenced by multiple interacting factors beyond duration alone.
\subsection{Topic Modeling and Thematic Consolidation of Videos}
\label{sec:videothemes}
To uncover latent thematic structures, we applied Latent Dirichlet Allocation (LDA)\cite{LDA2003} with optimized hyperparameters to our dataset of 13,215 videos. The model was trained on a combination of the \texttt{video\_description}, \texttt{hashtag\_names}, and \texttt{voice\_to\_text} fields. Preprocessing included stopword removal, lemmatization, and filtering of general terms.

The LDA model initially identified eight semantically related topics. The eight initial topics (along with the top five keywords for each) were:
\begin{sloppypar}
\begin{itemize}
    \item \textit{Topic 0 -- Digital Entrepreneurship \& Online Income:} 
    {entrepreneur}, {digital\_products}, {howtomakemoneyonline},
    {passive\_income}, {digitalmarketingtips}
   
    \item \textit{Topic 1 -- Property Investment \& Wealth Building:} 
    {investment}, property\_investing, {propertyportfolio}, 
    {moneytips}, {wealthbuilding}
    
    \item \textit{Topic 2 -- Forex \& Crypto Trading Education:} 
    {forex\_trading}, {trading}, {tradingforex}, 
    {investment}, {crypto\_trading}
    
    \item \textit{Topic 3 -- Property Buying \& Real Estate Tips:} 
    property\_investing, {buytolet}, {residential}, 
    {deposit}, {isa}
    
    \item \textit{Topic 4 -- Success Mindset \& Startup Motivation:} 
    {mindset}, {liquidity}, {successtips}, 
    {startupstories}, {entrepreneurialjourney}
    
    \item \textit{Topic 5 -- Budgeting \& Personal Finance:} 
    {saving}, {cashstuffingenvelopes}, {debtfree}, 
    {loudbudgeting}, {pennysavingchallenge}
    
    \item \textit{Topic 6 -- Side Hustles \& Work-from-Home:} 
    {sidehustleforbeginners}, {wfh}, {reseller}, 
    {sideincome}, {savemoney}
    
    \item \textit{Topic 7 – Crypto \& Beginner Investing:} 
    {btc}, {crypto}, {crypto\_trading}, 
    {xrp}, {education}
\end{itemize}
\end{sloppypar}
To improve interpretability and reduce overlap between closely related themes, these topics were consolidated into four broader thematic categories:
\begin{itemize}
    \item \textit{Entrepreneurship \& Side Hustles}: Topics 0, 4, 6
    \item \textit{Property Investing}: Topics 1, 3
    \item \textit{Active Trading (Forex \& Crypto)}: Topics 2, 7
    \item \textit{Saving \& Budgeting}: Topic 5
\end{itemize}

To further evaluate topic consistency, we also conducted a separate topic analysis using a BERT-based topic modeling approach. The identified themes were highly similar to the four merged categories adopted in this study.

Each video was then assigned to its most relevant LDA topic based on keyword overlap within the tokenized text fields. These assignments were subsequently mapped to the merged thematic categories for final analysis.

The topic categorization of the 13,215 analyzed videos is presented in Table~\ref{tab:topicstats}.

\begin{table}[tbh]
\centering
\small
\begin{tabular}{|l|c|}
\hline
\multicolumn{1}{|c|}{\textbf{Topic}} & \textbf{Number of Videos} \\
\hline
Entrepreneurship \& Side Hustles & 6447 \\
Property Investing & 1670 \\
Active Trading (Forex \& Crypto) & 3458 \\
Saving \& Budgeting & 1640 \\
\hline
\end{tabular}
\caption{Categorization of the 13,215 videos based on LDA topic assignment.}
\label{tab:topicstats}
\end{table}
\subsection{Analysis of Disclaimers in Videos}
\label{sec:disclaim}
This section primarily addresses RQ1 and RQ2 by examining the prevalence, characteristics, and engagement-related patterns of disclaimer usage.

To differentiate between finfluencers who clearly encourage their followers to conduct their own research, take independent action, or include other appropriate disclaimers, we analysed the videos for the presence of such disclaimers.

We analysed the fields \texttt{video\_description}, \texttt{voice\_to\_text}, and \texttt{hashtag\_names} for keywords or hashtags associated with disclaimers. A video was classified as containing a disclaimer if any of these fields contained either a keyword or a hashtag associated with disclaimers.
We emphasize that the disclaimer analysis was based on the aforementioned fields, and we did not examine the videos for visual or voice disclaimers. Therefore, some videos lacking \texttt{voice\_to\_text} may still contain disclaimers.

Disclaimers were identified using a rule-based text analysis applied to the aforementioned fields. This multi-source approach captures both formal disclaimers and informal disclaimer-related expressions embedded in social media content. The primary keyword dictionaries used in the analysis are summarized in Table~\ref{tab:disclaimerkeywords}.

A three-category classification framework was developed:
\begin{itemize}
    \item \textbf{Explicit disclaimers} are statements denying the provision of financial advice or encouraging independent verification (e.g., ``not financial advice'', ``DYOR''). See Table~\ref{tab:disclaimerkeywords} for the complete keyword dictionary.

    \item \textbf{Weak (implicit) disclaimers} are indirect responsibility-limiting statements that shift decision-making to the viewer without explicitly denying financial advice (e.g., ``invest at your own risk'', ``for educational purposes only''). Context-sensitive weak keywords were only retained when nearby advisory or responsibility-related terms were present.

See Table~\ref{tab:disclaimerkeywords} for the complete keyword dictionary.

    \item \textbf{Risk language} refers to statements describing financial uncertainty without limiting responsibility (e.g., ``trading involves risk'' and ``high risk''). These were treated separately from disclaimers.
\end{itemize}
The finance-context filter additionally incorporated trading-, pension-, and property-related terminology.

\begin{table}[tbh]
\centering
\footnotesize
\begin{tabular}
{|M{2.2cm}| p{5.3cm} |p{5.3cm}|}
\hline
\textbf{Category} & \multicolumn{1}{|c|}{\textbf{Text Keywords}} & \multicolumn{1}{|c|}{\textbf{Hashtag Keywords}} \\
\hline
Explicit &
not financial advice; this is not financial advice; not investment advice; not trading advice; not a recommendation; do your own research; consult a financial advisor; i am not a financial advisor; nfa; dyor
&
\#nfa; \#dyor; \#notfinancialadvice; \#notinvestmentadvice; \#notadvice; \#notarecommendation; \#cryptotipsnotadvice; \#stocktipsnotadvice; \#cryptonfa; \#tradingnfa
\\
\hline
Weak &
at your own risk; invest at your own risk; only invest what you can afford to lose; not responsible for losses; no liability accepted; results not guaranteed; for educational purposes only; you are responsible for your own decisions; take responsibility for your own decisions; for entertainment purposes only; for entertainment only; educational only; not responsible for any losses; no responsibility for losses
&
\#investatyourownrisk; \#entertainmentpurposesonly; \#educationalpurposesonly; \#tradingdisclaimer; \#pleaseconsult; \#onlyyoucanrisk; \#notliable
\\
\hline
Context-sensitive weak &
seek financial advice; seek advice
&

\\
\hline
Risk &
capital at risk; trading involves risk; risk involved; high risk; risky; market is volatile; this is risky
&

\\
\hline
\end{tabular}
\caption{Keyword dictionaries used for disclaimer classification.}
\label{tab:disclaimerkeywords}
\end{table}

Text was pre-processed through lowercasing, punctuation removal, and contraction standardisation to improve matching consistency. Disclaimer detection was implemented using keyword-based matching with regular expressions. For ambiguous expressions (e.g. ``seek advice''), a context-sensitive approach was applied, requiring nearby advisory or responsibility-related terms (e.g. ``consult'', ``advisor'', ``responsible'') within a defined textual window.

Disclaimer hashtags were identified both within hashtag fields and embedded video-description text. 

To reduce false positives, exclusion rules were applied to filter out general financial discussion (e.g. ``risk management'', ``portfolio'', ``stock market'') where risk-related terms do not function as disclaimers.
Because the follow-up dataset was collected from all videos posted by the original creator cohort rather than through finance-specific keyword retrieval, some videos in the later observation period were unrelated to finance. To reduce false positives in disclaimer analysis, we additionally applied a finance-context filter based on finance-related keywords, hashtags, and trading or investment terminology extracted from video descriptions, transcripts, and hashtags. Table \ref{tab:videodisclaimer} therefore reports both total detections and finance-context detections.

\paragraph{\bf Validation.} The classification approach was evaluated through manual review of a stratified random sample of 100 videos from each month, spanning explicit disclaimers, weak disclaimers, risk-only videos, and videos with no detected disclaimer. Manual inspection indicated a high level of agreement between the automated classifications and the human assessment, providing confidence that the rule-based framework achieves high precision for explicit disclaimer detection while substantially reducing false positives through contextual filtering and exclusion rules. Nevertheless, implicit or highly contextual disclaimer expressions remain inherently difficult to detect using rule-based approaches and may require machine-learning methods in future work.

\begin{table}[htb]
\centering
\small
\setlength{\tabcolsep}{9pt}
\begin{tabular}{|c|cc|cc|cc|cc|c|}
\hline

\textbf{M/Y} &
\multicolumn{2}{c|}{\textbf{Videos}} &
\multicolumn{2}{c|}{\textbf{Exp.}} &
\multicolumn{2}{c|}{\textbf{Weak}} &
\multicolumn{2}{c|}{\textbf{Risk}} &
\textbf{Users} \\ \cline{2-9}

&
\textbf{T} & \textbf{F} &
\textbf{T} & \textbf{F} &
\textbf{T} & \textbf{F} &
\textbf{T} & \textbf{F} &
\\ \hline
Apr-24 & 2203 & 2203 & 12 & 12 & 1 & 1 & 2 & 2 & 5 \\ \hline
May-24 & 2475 & 2475 & 31 & 31 & 1 & 1 & 7 & 7 & 15 \\ \hline
Jun-24 & 2023 & 2023 & 27 & 27 & 0 & 0 & 12 & 12 & 14 \\ \hline
Jul-24 & 2205 & 2205 & 40 & 40 & 1 & 1 & 10 & 10 & 15 \\ \hline
Aug-24 & 2301 & 2301 & 20 & 20 & 1 & 1 & 9 & 9 & 11 \\ \hline
Sep-24 & 2008 & 2008 & 41 & 41 & 0 & 0 & 9 & 9 & 9 \\ \hline

Oct-25 & 1714 & 684 & 7 & 7 & 0 & 0 & 10 & 10 & 9 \\ \hline
Nov-25 & 1450 & 589 & 12 & 12 & 3 & 3 & 20 & 17 & 9 \\ \hline
Dec-25 & 1762 & 742 & 15 & 15 & 0 & 0 & 11 & 11 & 10 \\ \hline
Jan-26 & 1844 & 817 & 5 & 5 & 1 & 1 & 25 & 23 & 10 \\ \hline
Feb-26 & 1633 & 665 & 10 & 9 & 39 & 20 & 19 & 19 & 13 \\ \hline
Mar-26 & 1876 & 703 & 15 & 15 & 35 & 2 & 9 & 9 & 12 \\ \hline
\end{tabular}
\caption{
Monthly disclaimer statistics across the two observation periods. T and F denote detections before and after finance-context filtering, respectively. Users denotes the number of distinct finfluencers posting finance-context videos containing disclaimer-related language.}
\label{tab:videodisclaimer}
\end{table}

Disclaimer-related language varied across both observation periods. Explicit disclaimers remained relatively infrequent overall, while the follow-up period showed noticeably higher counts of weak disclaimers and general risk-related language compared with the April–September 2024 observation window.

\begin{table}[tbh]
\centering
\small
\begin{tabular}{|c|p{11cm}|}
\hline
\textbf{M/Y} & \multicolumn{1}{|c|}{\textbf{Top 5 Hashtags \& Their Occurrence}} \\ \hline
Apr-24 & investingtips (7); investing101, investingforbeginners, investingexplained, crypto, bitcoin (4); cryptonews, cryptotok (3); ai, ripple, xrp, bitcoinnews, notfinancialadvice (2) \\

May-24 & 
 crypto (23); bitcoin (11); xrp (9); notfinancialadvice (8); investingforbeginners (7); turbo, investing, ripple, money, memecoin (6) \\
Jun-24& crypto (14); cryptocurrency (8); xrp, investing, investing101 (7)\\

Jul-24& cryptocurrency (15); personalfinance, investing (13); crypto (12); notfinancialadvice, bitcoin (11) \\
Aug-24 & cryptocurrency (17); bitcoin (16); investing (14); crypto (10); notfinancialadvice (9)  \\
Sep-24& bitcoin (32); cryptocurrency (24); notfinancialadvice (23); trading (18); crypto, investingforbeginners (17)  \\

Oct-25 & etf, indexfund (3); stockmarket, stocksandshares, investwithconfidence, yesiinvest (2) \\

Nov-25 &  personalfinance (4); femaleentrepreneur, abundancemindset, spiritualentrepreneur, budgeting, budget, tesla, trading, advice, retirement, mentor, invest, investing, sipp, nvidia, pension, exxonmobil, hedgefunds, advicetok, fundsmith, gold, moneytiktok, fintiktok, stockmarketinvesting (2) \\
Dec-25 &  pension (8); personalfinance, sipp, finance, nvidia, trading, investing (5) \\

Jan-26 & investing, stockmarket, investingtips (5); ai (3); trading, bitcoin, propertyinvesting, pension, finance, aiinvestments, landlord, hmo (2) \\
Feb-26 &  personalfinanceuk (9); pension (7); investing101, investingmadesimple, crypto, finance (5) \\ 
Mar-26 & personalfinance, taxyear, capcut (4); isaallowance, stockmarket (3) \\
\hline
\end{tabular}
\caption{Top 5 hashtags and their occurrences in videos with disclaimers.}
\label{tab:hashtag_occurrences}
\end{table}

Table~\ref{tab:hashtag_occurrences} summarizes the most frequent hashtags in disclaimer-related videos. Hashtags were lowercased, grouped by semantically equivalent variants, and ranked monthly by frequency.

Across both observation periods, disclaimer-related hashtags were dominated by cryptocurrency and investment-related terms (e.g., \#crypto, \#bitcoin, \#cryptocurrency, \#investing, and \#trading), indicating that disclaimer-related language was concentrated primarily in trading-oriented content.

\begin{table}[tbh]
\centering
\begin{tabular}{|l|c|}
\hline
\multicolumn{1}{|c|}{\textbf{Topic}} & \textbf{Number of Videos with Disclaimers} \\
\hline
Entrepreneurship \& Side Hustles & 24 \\
Property Investing & 21 \\
Active Trading (Forex \& Crypto) & 162 \\
Saving \& Budgeting & 8 \\
\hline
\end{tabular}
\caption{Distribution of the 215 videos containing disclaimer-related language across the four topic categories.}
\label{tab:disclaimbytopic}
\end{table}

\paragraph{\bf Disclaimers by Topic Category.}
This analysis addresses RQ3 by examining how disclaimer practices vary across thematic categories.

We examined how disclaimer-related videos were distributed across the four topic categories. Of the 215 videos containing disclaimer-related language, Table~\ref{tab:disclaimbytopic} shows the number assigned to each topic. Note that because topic modeling was conducted on the April--September 2024 dataset, this disclaimer-by-topic analysis is limited to the original observation period.

It is notable that \textit{Active Trading (Forex \& Crypto)} had the highest number of disclaimers, which may reflect the influence of recent guidance issued by regulatory bodies such as the Financial Conduct Authority (FCA).
\paragraph{\bf Analysis of Disclaimers by User.}
In the April--September 2024 dataset, 33 finfluencers used disclaimers in at least one video; among them, two also included profile-level disclaimers, three held verified accounts, and one account was no longer active.
\subsection{Sentiment Analysis of Comments}
\label{sec:sencom}
We have performed a sentiment analysis of the finfluencers' video comments for the April--September 2024 dataset. We used the Vader SentimentIntensityAnalyzer \cite{Hutto2014} to calculate sentiment scores for cleaned comments.
VADER was selected because it is computationally efficient and widely used for short-form social media text containing slang, emojis, and informal expressions.

To clean (pre-process) the comments, we removed URLs, mentions, and hashtags. We also normalized repeated characters and kept only alphabetic characters. We finally removed stop words using NLTK 
\cite{Loper2002}. 

Our sentiment analysis process involved calculating  a sentiment score for the text and a separate score for the emojis for each cleaned video's comment. A final sentiment score for each comment was computed as a weighted sum: $0.8 \times \text{text score} + 0.2 \times \text{emoji score}$. A higher weight was assigned to textual sentiment because emojis typically supplement rather than replace semantic meaning. This was repeated for all comments, and the average of the combined scores was used to determine the video-level sentiment.
  The video-level sentiment was classified as negative ($-1 \leq x < -0.05$), neutral ($-0.05 \leq x \leq 0.05$), or positive ($0.05 < x \leq 1$), where $x$ denotes the average sentiment score.
These thresholds follow commonly used polarity boundaries in VADER-based sentiment classification.

We also separately calculated the classification based solely on the text score in order to compare the two approaches.
The combined approach was adopted because emojis may also convey sentiment.

We present the summary of our analysis per month in Table \ref{tab:commentsentiment}. Note that the difference in the monthly video count here and in earlier sections is due to videos with no comments or videos whose comments are not accessible (e.g., they have been deleted). The figure next to each month indicates the number of videos that month with at least one accessible comment.

Most videos exhibited neutral or positive sentiment, with relatively few strongly negative videos. Disclaimer prevalence remained low across sentiment categories, with no consistent relationship between disclaimer usage and comment engagement. The most frequently occurring hashtags were broadly similar across the negative and neutral sentiment categories, with \#money, \#crypto, and \#forex appearing consistently in both. Overall, the findings highlight the limited use of disclaimers in financial TikTok content despite the prevalence of investment-related discussions.

\begin{table}[h!]
\begin{adjustwidth}{-1in}{-1in} 
\centering
\resizebox{1.3\textwidth}{!}{
\begin{tabular}{|l|c|c|c|c|c|l|}
\hline
\textbf{Mon.} & \textbf{Sentiment} & \textbf{\#V (C/T)}  &  \textbf{VDC}  &  \textbf{Avg. C.}&    \textbf{VD Avg. C.} & \multicolumn{1}{|c|}{\textbf{Top Hashtags}} \\
\hline
 Apr    & Negative& 102/118 & 1 & 3.1  & 1 &  money (35), forex (28),  crypto (28), investing (27), trading (22) \\
     (1465)                  & Neutral   & 313/319 &2 &   7.2 & 1.5 &  money (133), forex (85), trading (73), finance (59), crypto (57) \\
\hline May       & Negative & 110/129 & 5  & 2.4 & 2.4 & forex (35), money (34), trading (32), crypto (30), forextrading (23) \\
                (1609)      & Neutral   & 343/360  & 3 & 5.3 &  3.3 & money (106), crypto (79), forex (66), investing (58), trading (57)\\
\hline
 Jun      & Negative  & 70/91  & 3 & 2.4 & 4.3 &  money (26), crypto (21), forex (21), trading (19), wifimoney (18) \\
            (1339)           & Neutral   & 294/305 & 4 & 4.3  & 1.3 &  money (92), crypto (65), trading (51), investing (49), forex (46) \\
\hline
 Jul    & Negative  & 101/125 & 3 &  2.2  & 3.3 & trading (35), crypto (34), money (33),  forex (32), forextrading (26) \\
                  (1461)      & Neutral  & 329/342 & 9  & 5 & 3.3 &  money (124), crypto (79), investing (58), entrepreneur (58), trading (57)

 \\
\hline
Aug  & Negative & 77/97  & 2  & 3.2&  3 & money (31), trading (25), forex (23), crypto (22), forextrading (21) \\
            (1599)            & Neutral  & 250/259 & 3 &  9.9& 3 &  money (110), trading (42), investing (41), crypto (39), viral (38) \\
\hline
Sept & Negative & 39/52  & 2 &2.3 & 4.5 & money (20), investing (15), trading (12), forex (9), forextrading (7) \\
             (1368)         & Neutral  & 190/196 & 6 & 10.6 & 11.5 &  money (82), investing (52), trading (46), crypto (43), forex (41) \\
\hline
\end{tabular}
}
\caption{\scriptsize Summary of sentiment analysis of videos. The number of videos (\#V) is shown in the format (C/T), where C is the Combined score (i.e., Text and Emoji) and T is Text Only. Hashtag counts are based on the Combined score. \textbf{VDC} denotes the number of videos containing disclaimers. \textbf{Avg. C.} refers to the average number of comments per video, while \textbf{VD Avg. C.} is the average number of comments on videos that include a disclaimer.}
\label{tab:commentsentiment}
\end{adjustwidth}
\end{table}
\section{Conclusion}
\label{sec:summary}
This study provides a longitudinal analysis of UK TikTok finfluencers, examining engagement dynamics, disclosure practices, hashtag trends, sentiment patterns, and network structure across two observation periods. Topic modeling of the April--September 2024 dataset identified four dominant themes: \textit{Entrepreneurship}, \textit{Property Investing}, \textit{Active Trading}, and \textit{Saving \& Budgeting}, reflecting the core narratives in financial discourse on the platform.

Engagement levels varied substantially across the dataset, although video duration showed only a negligible association with engagement rate. Social network analysis indicated a collaborative ecosystem in which mid-tier finfluencers frequently acted as bridges between creator groups.

Sentiment across videos skewed neutral to positive, with few instances of negative tone. 
Explicit disclaimers remained relatively uncommon across both observation periods, although weak disclaimers and general risk-related language became more prevalent in parts of the follow-up dataset.
Notably, disclaimers were predominantly associated with active trading content (e.g., forex and cryptocurrency), likely reflecting recent regulatory guidance emphasizing the importance of risk disclosure in financial content. However, their presence showed limited association with engagement patterns, while financial hashtags such as \#crypto and \#bitcoin remained common in disclaimer-related content.

These findings suggest that transparency remains limited within the UK finfluencer ecosystem, with explicit disclaimers remaining relatively uncommon despite the prominence of trading-related content.

This study has several limitations. The analysis focuses on UK-based TikTok finfluencers identified through finance-related keywords and therefore may not generalize to other platforms, regions, or influencer categories. Some analyses, including topic modeling and sentiment analysis, were restricted to the original observation period due to data availability constraints. Videos lacking transcribed speech may also contain undetected disclaimers.

Overall, the findings highlight the need for stronger transparency and regulatory oversight of finfluencer content. 
Based on the low prevalence of explicit disclaimers, platforms and regulators should encourage clearer risk disclosures, particularly for trading-related content. Social media platforms could also implement standardized content labeling or financial literacy prompts to help users better assess the reliability of financial information.

\section*{Acknowledgement}
This work was supported in part by the REPHRAIN National Research Centre on Privacy, Harm Reduction and Adversarial Influence Online.

\begingroup
\bibliographystyle{splncs04}
\bibliography{references}
\endgroup

\end{document}